\documentclass[11pt]{article}
\usepackage{epsfig}
\usepackage{graphicx}
\usepackage{cite}

\newcommand{\BABARPubYear}    {04}

\newcommand{\BABARProcNumber} {177}
\newcommand{\SLACPubNumber} {11046}

\newcommand{\babar}{\textsc{BaBar}}
\newcommand{\epem}{\ensuremath{e^+e^-}}
\newcommand{\fbinv}{\ensuremath{\mathrm{fb}^{-1}}}
\newcommand{\MeVcc}{\ensuremath{\mathrm{MeV}/c^2}}
\newcommand{\GeVcc}{\ensuremath{\mathrm{GeV}/c^2}}
\newcommand{\GeVc}{\ensuremath{\mathrm{GeV}/c}}
\newcommand{\csbar}{c\overline{s}}
\newcommand{\ccbar}{c\overline{c}}
\newcommand{\qqbar}{q\overline{q}}
\newcommand{\Jpsi}{J/\psi}
\newcommand{\pip}{\pi^+}
\newcommand{\pim}{\pi^-}
\newcommand{\piz}{\pi^0}
\newcommand{\BBbar}{\ensuremath{B\overline{B}}}
\newcommand{\Bp}{B^+}
\newcommand{\Bm}{B^-}
\newcommand{\Kp}{K^+}
\newcommand{\Km}{K^-}
\newcommand{\KS}{K^0_S}
\newcommand{\BF}{\mathcal{B}}
\newcommand{\Dz}{D^0}
\newcommand{\Dzb}{\overline{D}{}^0}
\newcommand{\Dstp}{D^{*+}}
\newcommand{\Dstarz}{D^{*0}}
\newcommand{\Dsp}{D_s^+}
\newcommand{\Dsstp}{D_s^{*+}}
\newcommand{\DsJ}{D_{sJ}}
\newcommand{\pbar}{\overline{p}}
\newcommand{\Kbar}{\overline{K}}
\newcommand{\Ds}{\Dsp}
\newcommand{\Dbar}{\overline{D}}
\newcommand{\Db}{\overline{D}}

\newcommand{\Dm}{D^-}
\newcommand{\Kstarzb}{\overline{K}{}^{*0}}
\newcommand{\Dss}{D_s^*}

\newcommand{\DsTT}{D_{sJ}^*(2317)^+}
\newcommand{\DsTO}{D_s^{*}(2112)^+}
\newcommand{\DsFE}{D_{sJ}(2460)^+}
\newcommand{\DsTS}{D_{s1}(2536)^+}

\long\def\inst#1{\par\nobreak\kern 4pt\nobreak
    {\it #1}\par\vskip 10pt plus 3pt minus 3pt}

\begin{document}
{\pagestyle{empty}

\begin{flushright}
SLAC-PUB-\SLACPubNumber \\
BABAR-PROC-\BABARPubYear/\BABARProcNumber \\
March 3, 2005 \\
\end{flushright}

\par\vskip 4cm

\begin{center}
\Large \bf
Beyond CP violation: hadronic physics at BaBar 
\end{center}
\bigskip

\begin{center}
\large 
Gabriella Sciolla \\
Massachusetts Institute of Technology, Department of Physics\\
Room 26-443, 77 Massachusetts Avenue, Cambridge, MA \\
(for the BaBar Collaboration)
\end{center}
\bigskip \bigskip

\begin{center}
\large \bf Abstract
\end{center}
I report on recent studies of hadronic physics performed 
by the \babar\ Collaboration. 
Emphasis is given to the measurement of the properties of 
newly discovered charmed hadrons and 
to the searches for light and heavy pentaquarks. 
\vfill
\begin{center}
Contributed to the Proceedings of the First APS Topical Group Meeting on
Hadron Physics\\ 
Fermilab, Batavia, IL (October 24-26, 2004). 
\end{center}

\vspace{1.0cm}
\begin{center}
{\em Stanford Linear Accelerator Center, Stanford University, 
Stanford, CA 94309} \\ \vspace{0.1cm}\hrule\vspace{0.1cm}
Work supported by Department of Energy contracts
DE-FC02-94ER40818 and DE-AC02-76SF00515.  
\end{center}

\section{Introduction}

The \babar\ experiment~\cite{detector} operates at the PEP-II $\epem$ asymmetric B-factory  at SLAC. 
The abundant data sample of over 240\,\fbinv\ to date, allows a rich and active research program.
The primary goal of the experiment is to perform a quantitative test 
of the Standard Model, which can be achieved by precise measurements
of CP violation in the $B$ system, and of the CKM matrix elements
related to the determination of the sides of the Unitarity Triangle.
The study of $B$ decays also provides a sensitive probe for New Physics
beyond the Standard Model and allows for tests of Factorization, HQET,
HQE and SCET. 

In addition to studying  $B$ decays,
\babar\ is also active in the area of
spectroscopy of lighter particles.
This article focuses on two such topics:  
charm spectroscopy and searches for pentaquarks. 


\section{Recent results in charm spectroscopy}

In 2003  \babar\  discovered a narrow resonance~\cite{D-ref1}
in the channel $\Dsp\piz$ with a mass of 2317\,\MeVcc.
The existence of this state, named $\DsTT$,
was soon confirmed by CLEO~\cite{D-ref2} and Belle~\cite{D-ref3}. 
Subsequently, the CLEO Collaboration discovered another $\Dsstp\piz$
narrow state~\cite{D-ref2}, the $\DsFE$,
promptly confirmed by \babar~\cite{D-ref3} and Belle~\cite{D-ref4}. 

The discovery of the $\DsJ$ resonances received a lot of attention by the theoretical 
community and revived the entire field of charm spectroscopy.
Many hypotheses were 
formulated to explain the unexpected 
narrow widths and masses of these states that are not in agreement with the theoretical predictions. 
These resonances are usually interpreted as $P$-wave $\csbar$ quark 
states~\cite{D-ref5,D-ref6,D-ref8,D-ref9}, although other 
interpretations~\cite{D-ref10,D-ref11,D-ref12,D-ref13,D-ref14} cannot be ruled out.   

In this section I will report on the  
improved measurements of the properties of the  
$\DsTT$ and $\DsFE$~\cite{D,D2}.   
I will also review the 
recently published evidence for the $X(3872)$~\cite{X} and the status of the search 
for the $\DsJ(2632)^+$~\cite{D3}.

\subsection{Mass of the $D^*_{sJ}(2317)^+$ and $D_{sJ}(2460)^+$ in $\epem\to\ccbar$}
\babar\ recently published~\cite{D2} new measurements for the masses of the 
$\DsTT$ and $\DsFE$.
This analysis, based on a dataset of 125 fb$^{-1}$, 
reconstructs the $\DsJ$ resonances produced in $\ccbar$ events 
in four final states: 
$\Dsp\piz$, $\Dsp\gamma$, $\Dsp\piz\gamma$, and
$\Dsp\pip\pim$.
The $\Dsp$ candidates are reconstructed in $K^+K^-\pi^+$ through 
$\Kbar{}^{*0}K^+$ and $\phi\pip$ intermediate states.
The  $\DsJ$ candidates are obtained 
by forming invariant mass combinations of 
the reconstructed $\Dsp$ mesons with $\pi^0$, $\gamma$, and $\pi^\pm$ particles.  

Figure~\ref{D2-fig2+4} (left) shows the invariant mass of the $\DsTT$
candidates reconstructed in 
the $\Dsp\piz$ final state. A clear $\DsTT$ signal is visible on top of 
the combinatorial background. Reflections from 
$\DsFE\to\Dsp\piz\gamma$ and 
$\DsTO\to\Dsp\gamma$ are shown in dark and light gray, respectively. 
\begin{figure}\centering
  \includegraphics[width=7.5cm,height=7.7cm]{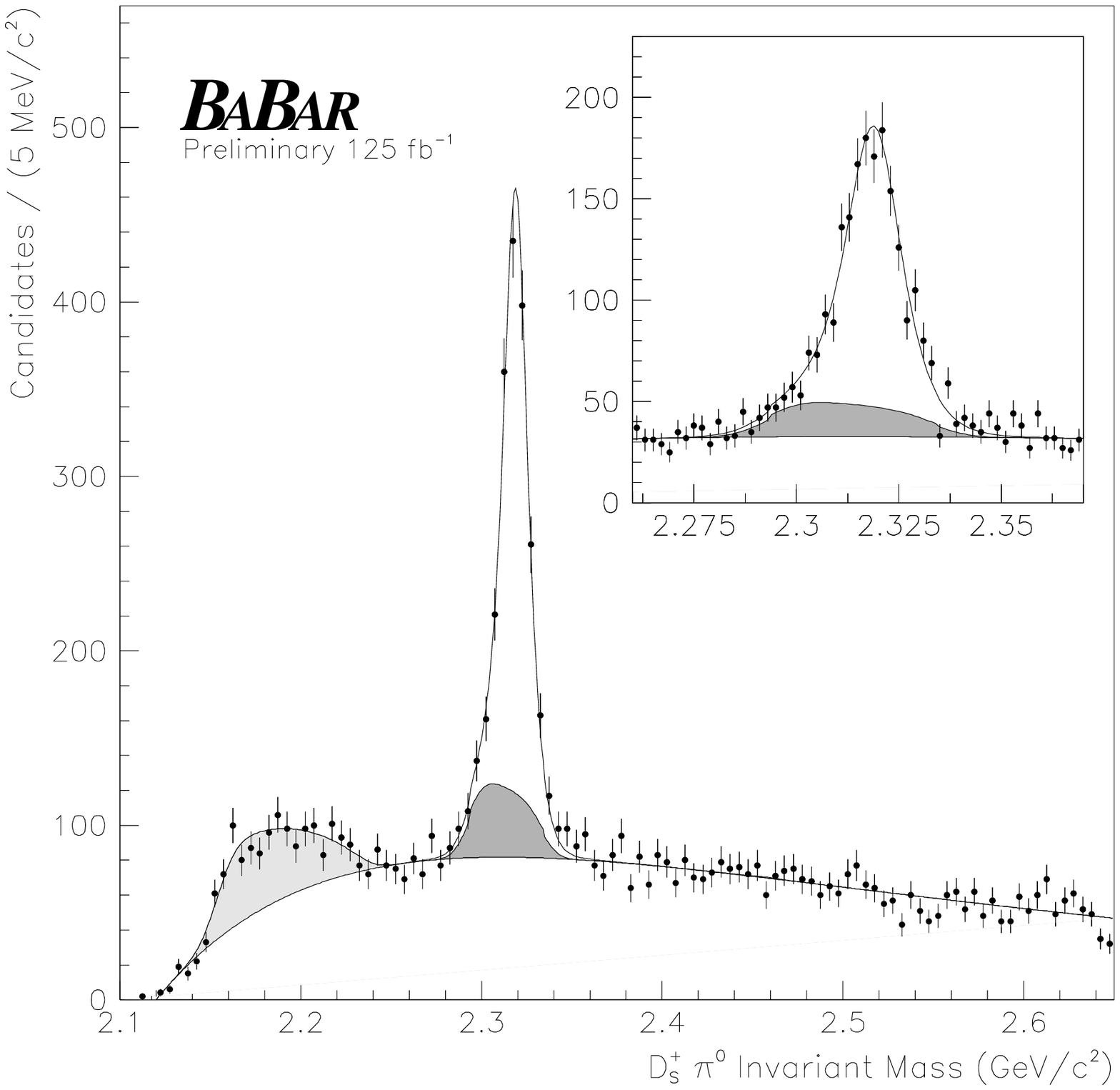}
  \includegraphics[width=7.5cm,height=7.7cm]{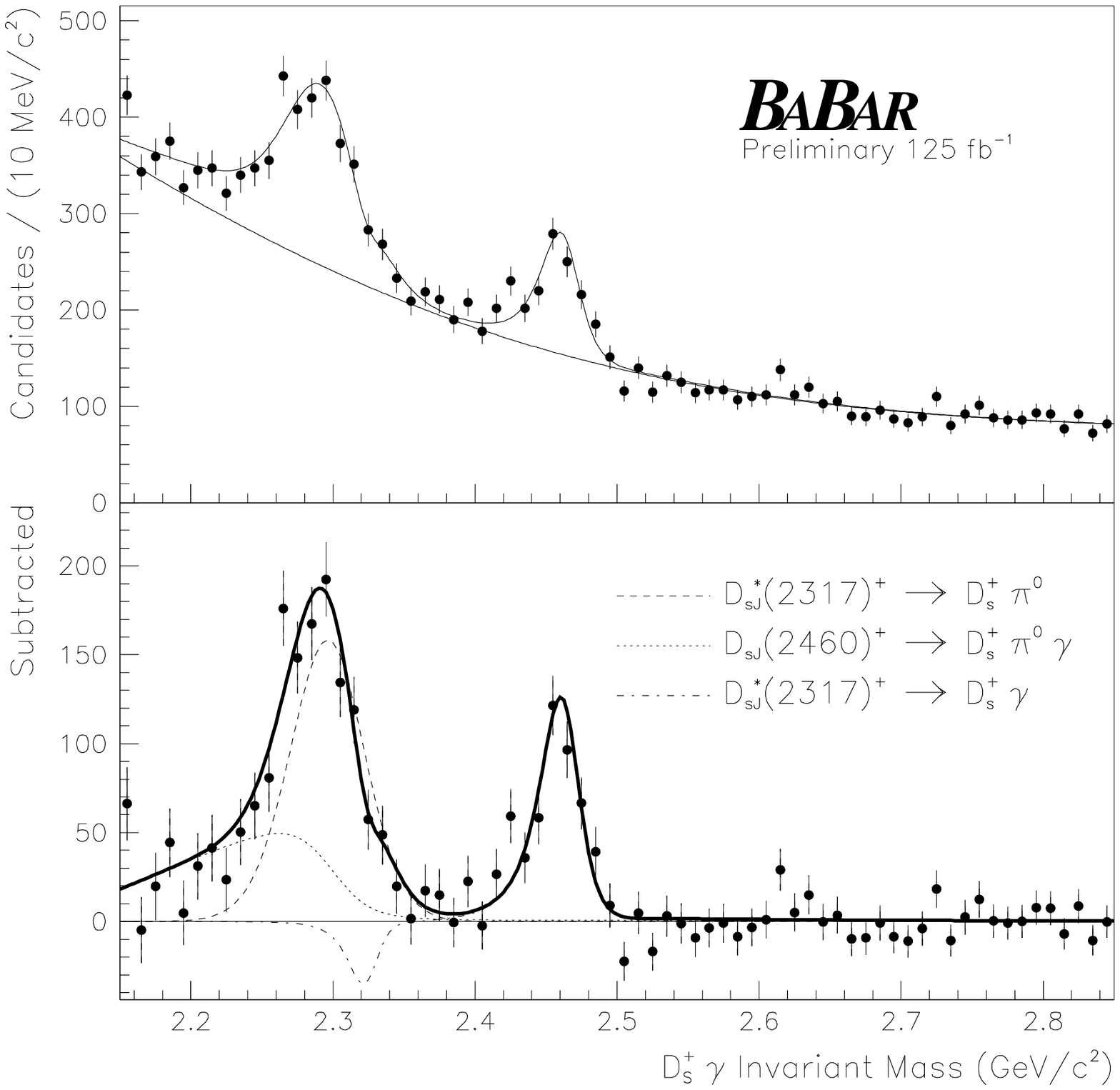}
  \caption{Left: $\Ds\piz$ invariant mass distribution.
    The solid curve is the result of an unbinned
    likelihood fit. The dark (light) gray region is the contribution from the
    $\DsFE\to\Ds\piz\gamma$ ($\DsTO\to\Ds\gamma$) reflection.
    The inset is an expanded view near the $\DsTT$ mass.
    Right: 
    $\Ds\gamma$ invariant
    mass distribution before (top) and after (bottom) subtraction of the 
    combinatorial background. 
    The solid curve is the result of an unbinned
    likelihood fit. 
    Various contributions to the signal and reflection portions of the
    fit are overlaid (dashed lines). 
  }
  \label{D2-fig2+4}
\end{figure}
A likelihood fit measures  
a signal yield of  $1275\pm45$ and an invariant mass of 
\[
m(\DsTT) = (2318.9\pm0.3\;({\rm stat.})\pm0.9\;({\rm syst.}))\,\MeVcc .
\]

Figure~\ref{D2-fig2+4} (right) shows the  $\DsFE$ invariant mass 
in the  $\Dsp\gamma$ final state. 
The signal is clearly visible on top of a falling combinatorial background. 
The broader peak centered around $2.3\,\GeVcc$ is a combination of two reflections.  
The larger reflection is due to 
$\Ds$ mesons from the decay $\DsTT\to\Ds\piz$ combined
with one of the photons from the $\piz$;  the other 
is produced in a similar fashion from $\DsFE\to\Ds\piz\gamma$ decay.
A signal of $509 \pm 46$ $\DsFE$ candidates allows us to measure
\[
m(\DsFE\to\Ds\gamma) = (2457.2\pm 1.6\;({\rm stat.}) \pm 1.3\;({\rm syst.}))\,\MeVcc.
\]

Figure~\ref{D2-fig6} (left) illustrates the invariant mass distribution for  
the $\DsFE$ decaying into $\Ds\piz\gamma$.  
This three body decay is dominated by the intermediate state 
$\DsTO\piz$. 
\begin{figure}\centering
  \includegraphics[width=7cm]{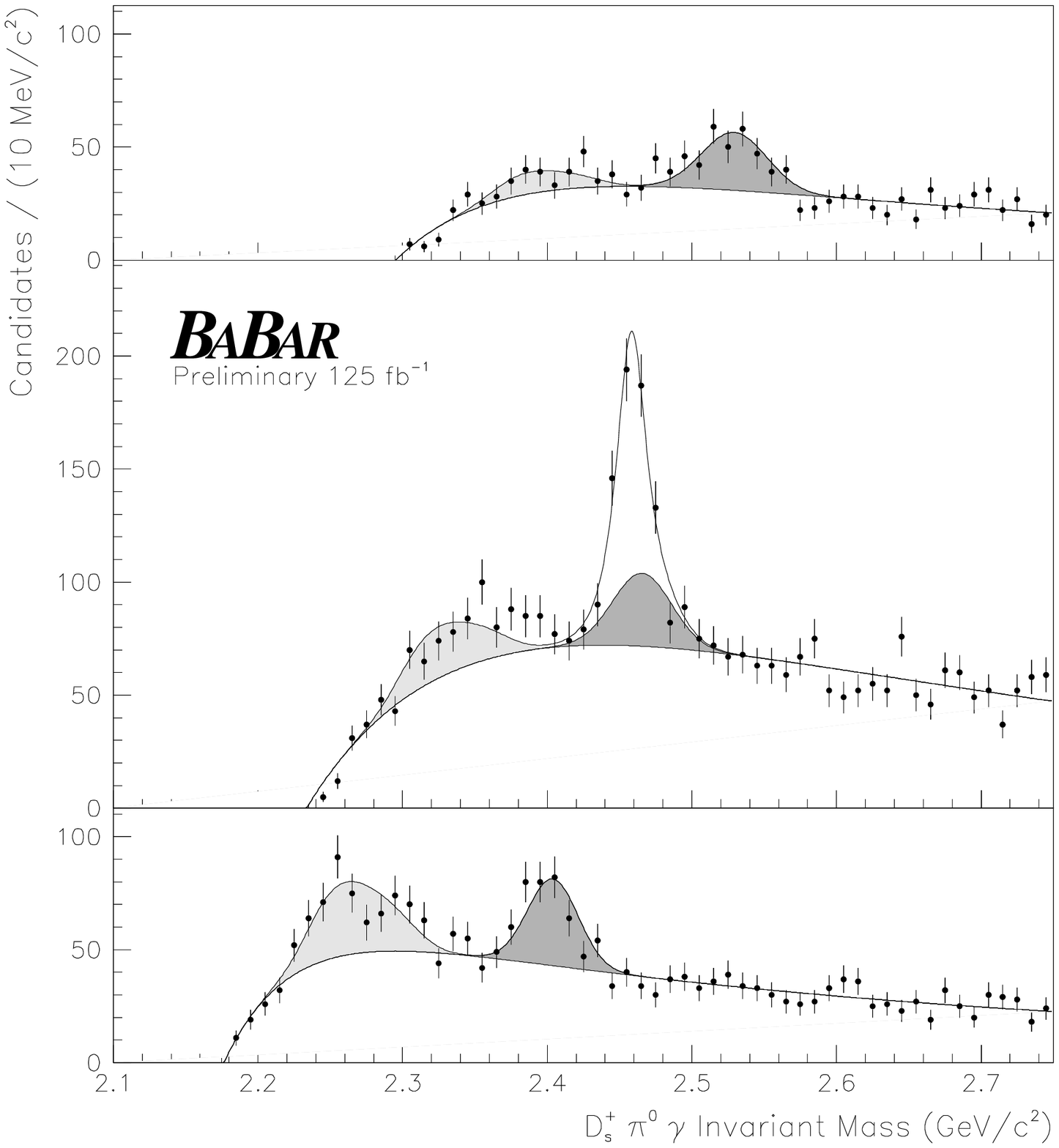}
  \includegraphics[width=8.5cm,height=7cm]{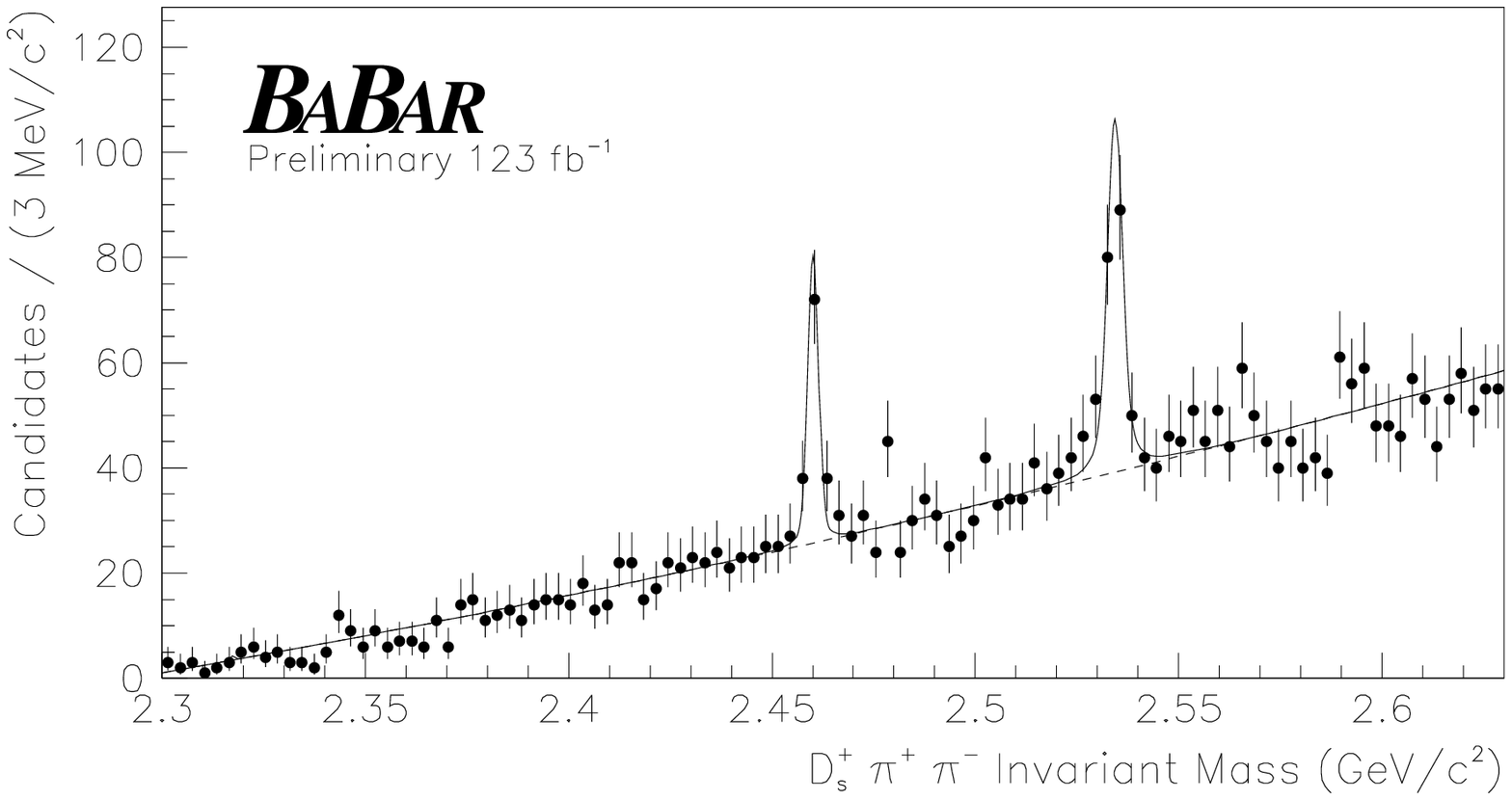}
  \caption{Left: $\Ds\piz\gamma$ mass spectrum
    of the events that fall in the $\Ds\gamma$ signal region (center
    plot), the $\Ds\gamma$ high mass sideband (top plot), and the
    low mass sideband (bottom plot). 
    Reflections from (dark gray)
    $\DsTT\to\Ds\piz$ and (light gray) $\DsTO\to\Ds\gamma$ decays
    appear well modeled in the Monte Carlo.
    Right: $\Ds\pip\pim$ invariant
    mass distribution for candidates that satisfy the requirements
    discussed in the text. The solid curve is the result of an unbinned
    likelihood fit.}
  \label{D2-fig6}
\end{figure}
A likelihood fit yields a signal of $292\pm29$ $\DsFE$ events. 
Two reflections are visible in the distribution: 
the light gray shoulder is  due to $\DsTO\to\Ds\gamma$ decays;
the dark gray peaking background is due to $\DsTT\to\Ds\piz$ decays combined
with unassociated photons. 
The position and shape of the reflections is extracted  from  Monte Carlo 
simulation
and validated on the data side bands as shown in figure~\ref{D2-fig6} left-top and 
left-bottom plots. 
The mass of the  $\DsFE$ extracted from this measurement is 
\[
m(\DsFE\to\Ds\piz\gamma) = (2459.1\pm 1.3\;({\rm stat.})\pm 1.2\;({\rm syst.}))\,\MeVcc.
\]

Figure~\ref{D2-fig6} (right) shows the invariant mass of the  $\DsFE$  candidates 
reconstructed in the $\Ds\pip\pim$  final state. 
A signal of $67\pm11$ $\DsFE$ decays is clearly visible on top of 
the combinatorial background; a second peak due to $124\pm 18$ 
$\DsTS$ particles is also evident. 
The masses for these two particles are extracted by using a likelihood fit:
\begin{eqnarray*}
m(\DsFE\to\Ds\pip\pim) &=& (2460.1\pm 0.3\;({\rm stat.})
          \pm 1.2\;({\rm syst.}))\,\MeVcc,\\
m(\DsTS) &=& (2534.3\pm 0.4\;({\rm stat.})
          \pm 1.2\;({\rm syst.}))\,\MeVcc.
\end{eqnarray*}

The mass measurements mentioned above are combined to
obtain the following average:
\begin{eqnarray*}
m(\DsFE) &=& (2459.4\pm 0.3\;({\rm stat.}) \pm 1.0\;({\rm syst.}))\,\MeVcc.
\end{eqnarray*}

By combining the efficiency corrected yields,
the following relative branching fractions are measured: 
\begin{eqnarray*}
\frac{\BF(\DsFE\to\Ds\gamma)}{\BF(\DsFE\to\Ds\piz\gamma)} &=&
0.375 \pm 0.054\;({\rm stat.}) \pm 0.057\;({\rm syst.}),\\
\frac{\BF(\DsFE\to\Ds\pi^+\pi^-)}{\BF(\DsFE\to\Ds\piz\gamma)} &=&
0.082 \pm 0.018\;({\rm stat.}) \pm 0.011\;({\rm syst.}).
\end{eqnarray*}

\subsection{Properties of $\DsTT$ and $\DsFE$ from exclusive $B$ decays}
Pure samples of $\DsTT$ and $\DsFE$ can be obtained by exclusive 
reconstruction of  
the decay $B\to D^{(*)+}_{sJ}\overline{D}^{(*)}$~\cite{D}. 
This analysis, performed on a sample of 113\,\fbinv, measures 
the spin and branching fractions of the $\DsJ$ resonances.

The $\Dbar$ and $\Dsp$ mesons are reconstructed in the following decay modes:
 $\Dzb\to\Kp\pim$,
 $\Kp\pim\piz$, $\Kp\pim\pip\pim$; $\Dm\to\Kp\pim\pim$; 
$\Ds\to\phi\pip$ ($\phi\to\Kp\Km$), $\Kstarzb\Kp$ ($\Kstarzb\to\Km
\pip$).
The $D^*$ candidates are reconstructed in the decay modes $D^{*+}\to \Dz\pip$,
 $\Dstarz\to \Dz\piz$, $\Dz\gamma$, and $\Dss\to\Dsp\gamma$. 
The selected pairs of $D_s^{(*)+}$ and $\overline{D}^{(*)}$
candidates are combined with a photon  or a $\piz$ to complete 
the  $B$ decay. 
Cuts on the energy and invariant mass of the $B$ candidate are used to enhance the 
purity of the signal. 

Figure \ref{B-fig2} shows the $\DsJ$ invariant mass for the selected $B$ candidates 
for  12 different  $\Dbar{}^{(*)} \DsJ$ final states. 
The yields, the corresponding branching fractions and the statistical significance   
of each channel are listed in Table~\ref{tab:br1st}.
\begin{figure}
  \begin{center}
    \hspace*{-0mm}\includegraphics[width=5.5cm,height=6cm]{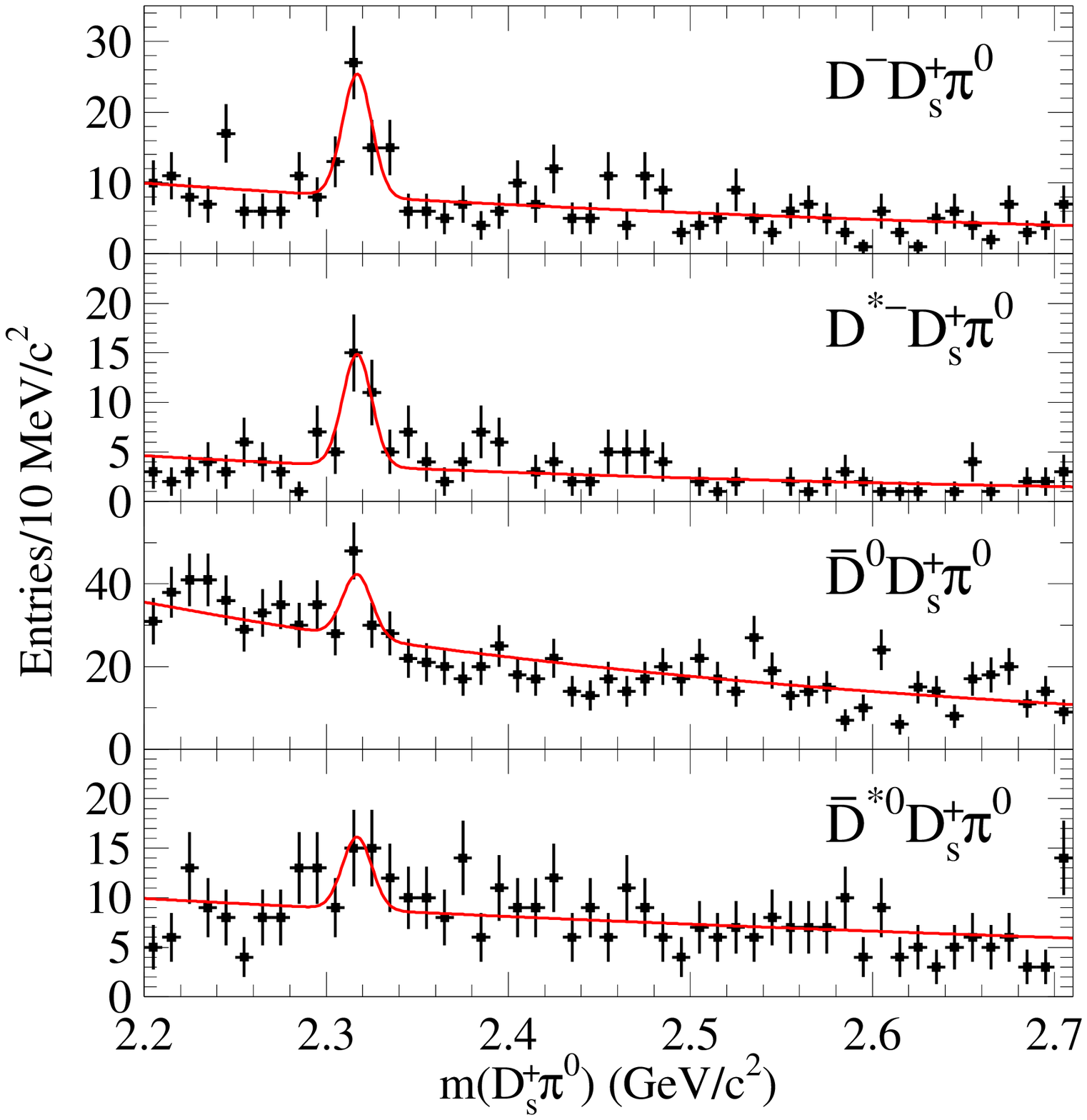}
    \hspace*{-5mm}\includegraphics[width=5.5cm,height=6cm]{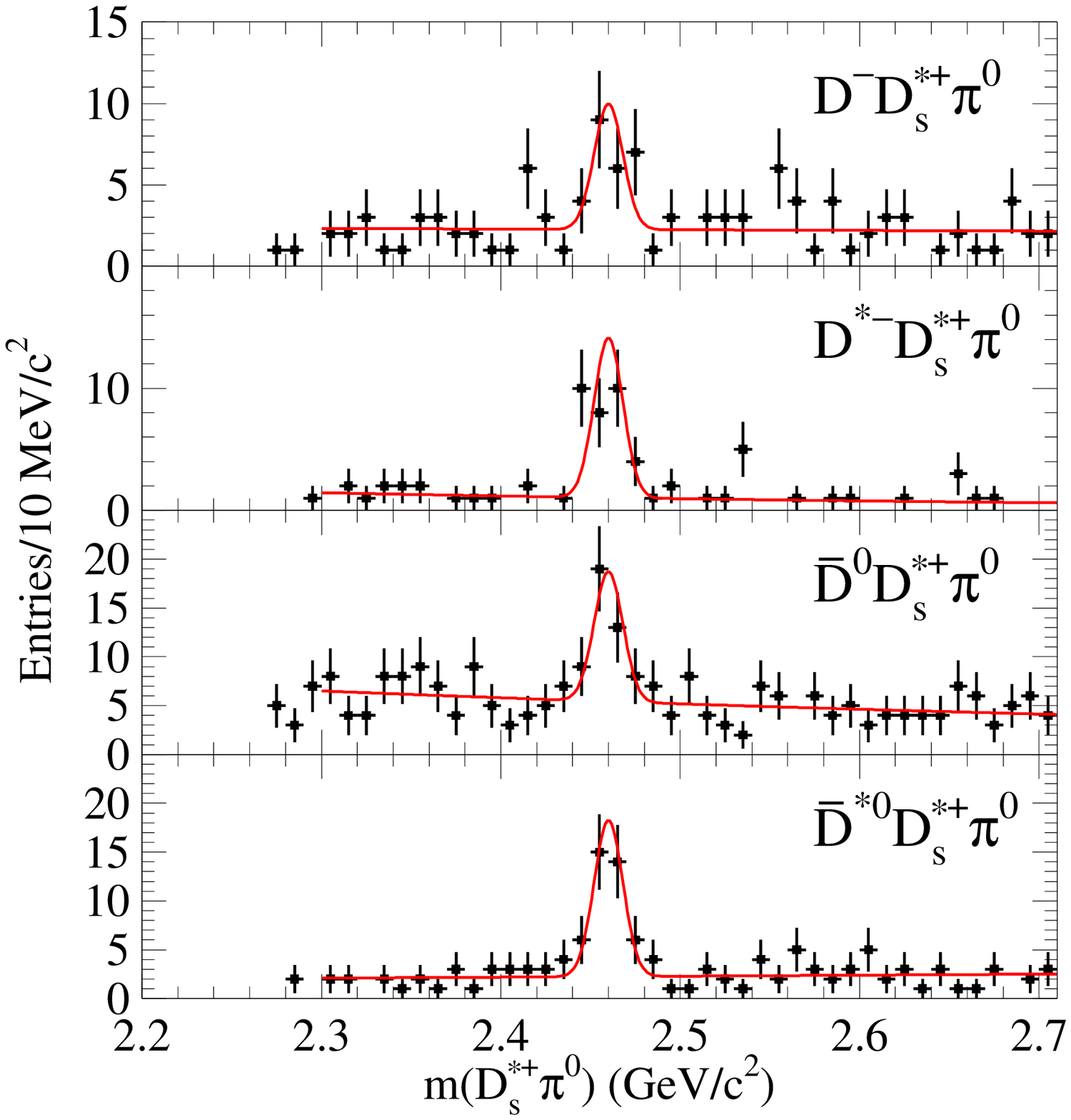}
    \hspace*{-5mm}\includegraphics[width=5.5cm,height=6cm]{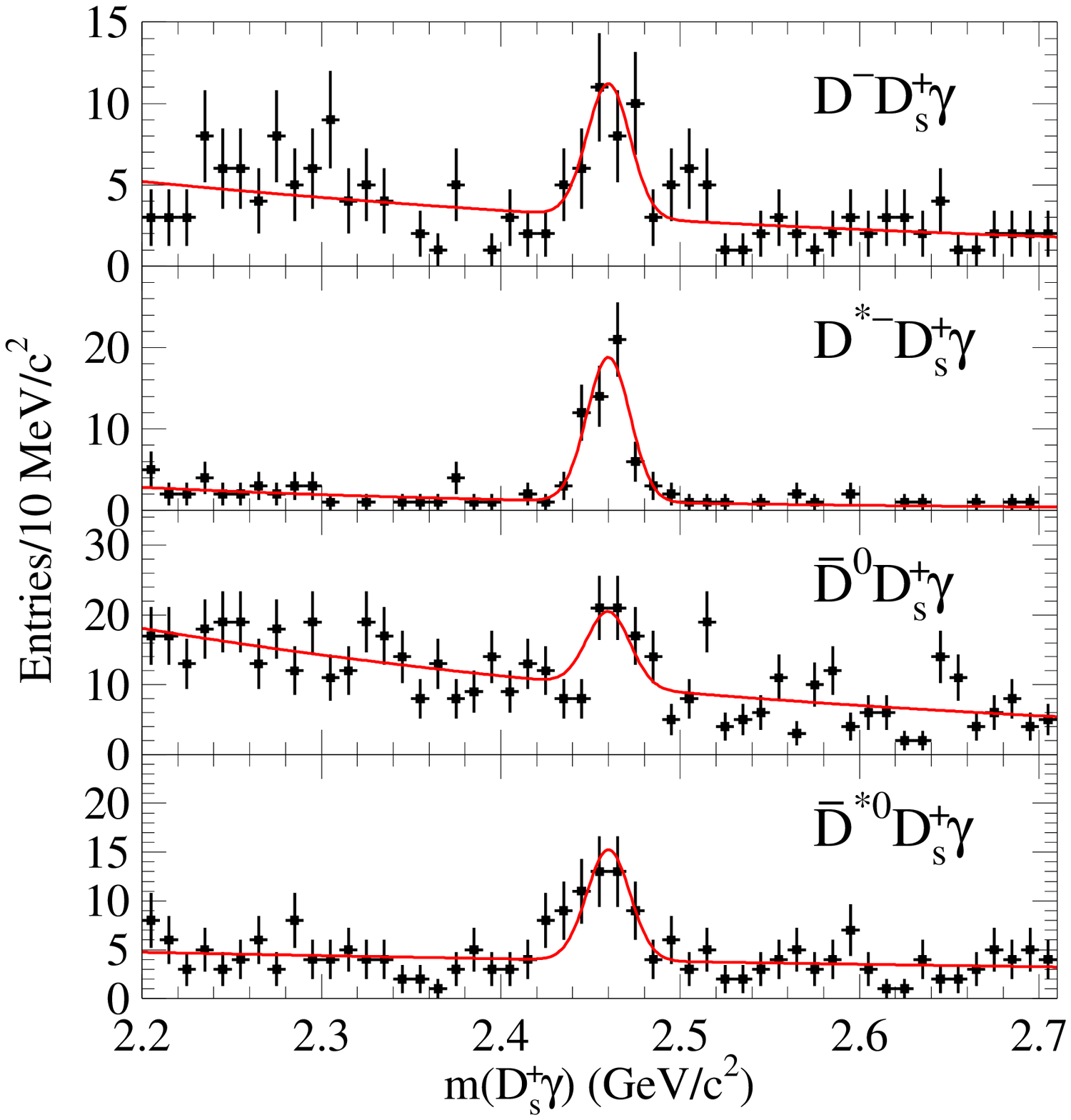}
  \end{center}
  \caption{ $D_s^+\piz$ (left), $D_s^{*+}\piz$ (center), and
    $D_s^+\gamma$ (right) mass spectra of the selected $B$ signal candidates
    for the 12 $\Dbar{}^{(*)} \DsJ$ final states. Curves are the results
    of the fits.}
  \label{B-fig2}
\end{figure}
\begin{table}
  \centering
  \caption{Event yields, final branching fractions ${\mathcal B}$,
    and significance for $B\to\DsJ^{(*)+}\Dbar{}^{(*)}$ decays.
    The first error on ${\cal B}$ is statistical, the second is systematic,
    and the third is from the $\Db$ and $\Ds$ branching
    fractions.} \label{tab:br1st}
\vspace{0.3cm}
  \begin{tabular}{l@{$\rightarrow$}llr@{$\,\pm\,$}r@{.}lcr@{.}l}
    \hline\hline
    \multicolumn{3}{c}{$B$ mode}
    & \multicolumn{3}{c}{Yield}  & ${\cal B}(10^{-3})$ & \multicolumn{2}{c}\
  {Significance} \\
    \hline
  \rule[0mm]{0mm}{1.1em}$B^0$&$ D_{sJ}^{*}(2317)^+D^{-}$ &[$D_{s}^{+} \piz$]
  & $34.7$&8&0 & $1.8 \pm 0.4\pm 0.3 ^{+0.6}_{-0.4}$ &\hspace*{5mm}  5&5 \\
  $B^0$&$ D_{sJ}^{*}(2317)^+D^{*-}$ &[$D_{s}^{+} \piz$]
  & $23.5$&6&1 & $1.5 \pm 0.4\pm 0.2 ^{+0.5}_{-0.3}$ &  5&2 \\
  $B^+$&$ D_{sJ}^{*}(2317)^+\overline{D}^{0}$ &[$D_{s}^{+}
  \piz$]
  & $32.7$&10&8 & $1.0 \pm 0.3\pm 0.1 ^{+0.4}_{-0.2}$ &  3&1 \\
  $B^+$&$ D_{sJ}^{*}(2317)^+\overline{D}^{*0}$ &[$D_{s}^{+}
  \piz$]
  & $17.6$&6&8 & $0.9 \pm 0.6\pm 0.2 ^{+0.3}_{-0.2}$ &  2&5 \\
  $B^0$&$ D_{sJ}(2460)^+D^{-}$ &[$D_{s}^{*+} \piz$]
  & $17.4$&5&1 & $2.8 \pm 0.8\pm 0.5 ^{+1.0}_{-0.6}$ &  4&2 \\
  $B^0$&$ D_{sJ}(2460)^+D^{*-}$ &[$D_{s}^{*+} \piz$]
  & $26.5$&5&7 & $5.5 \pm 1.2\pm 1.0 ^{+1.9}_{-1.2}$ &  7&4 \\
  $B^+$&$ D_{sJ}(2460)^+\overline{D}^{0}$ &[$D_{s}^{*+}
  \piz$]
  & $29.0$&6&8 & $2.7 \pm 0.7\pm 0.5 ^{+0.9}_{-0.6}$ &  5&1 \\
  $B^+$&$ D_{sJ}(2460)^+\overline{D}^{*0}$ &[$D_{s}^{*+}
  \piz$]
  & $30.5$&6&4 & $7.6 \pm 1.7\pm 1.8 ^{+2.6}_{-1.6}$ &  7&7 \\
  $B^0$&$ D_{sJ}(2460)^+D^{-}$ &[$D_{s}^{+} \gamma$]
  & $24.8$&6&5 & $0.8 \pm 0.2\pm 0.1 ^{+0.3}_{-0.2}$ &  5&0 \\
  $B^0$&$ D_{sJ}(2460)^+D^{*-}$ &[$D_{s}^{+} \gamma$]
  & $53.0$&7&8 & $2.3 \pm 0.3\pm 0.3 ^{+0.8}_{-0.5}$ & 11&7 \\
  $B^+$&$ D_{sJ}(2460)^+\overline{D}^{0}$ &[$D_{s}^{+}
  \gamma$]
  & $31.9$&9&0 & $0.6 \pm 0.2\pm 0.1 ^{+0.2}_{-0.1}$ &  4&3 \\
  \rule[-0.5em]{0mm}{1.5em}$B^+$&$ D_{sJ}(2460)^+\overline{D}^{*0}$ &[$D_{s}^{+}
  \gamma$]
  & $34.6$&7&6 & $1.4 \pm 0.4\pm 0.3 ^{+0.5}_{-0.3}$ &  6&0 \\
    \hline\hline
  \end{tabular}
\end{table}

From the measured branching fractions for {$B \to D_{sJ}^+(2460)\overline{D}^{(*)}$} 
in  $D_s^{*+}\piz$ and in
$D_s^{+}\gamma$, we find the  ratio
\[
\frac{{\cal B}(D_{sJ}(2460)^+ \to D_s^{+}\gamma)}{{\cal B}(D_{sJ}(2460)^+ \to
  D_s^{*+}\piz)} = 0.274\pm 0.045 \pm 0.020,
\]
in agreement  with the prediction from~\cite{D-ref6}.

A helicity analysis is performed for the $\DsFE$ state, using the
decays $B^+\rightarrow \DsFE \overline{D}^0$ and
$B^0\rightarrow \DsFE D^- $, with $\DsFE \rightarrow D^+_s\gamma$.
The helicity angle $\theta_h$ is defined as the angle between the
$\DsJ$ momentum in the $B$-meson rest frame and the $D_s$ momentum
in the $\DsJ$ rest frame. 
Figure~\ref{fig:helicity} shows the resulting angular distribution,
corrected for detector acceptance and selection efficiency. 
\begin{figure}
 \begin{center}
  \hspace*{-5mm}\includegraphics[width=9.0cm]{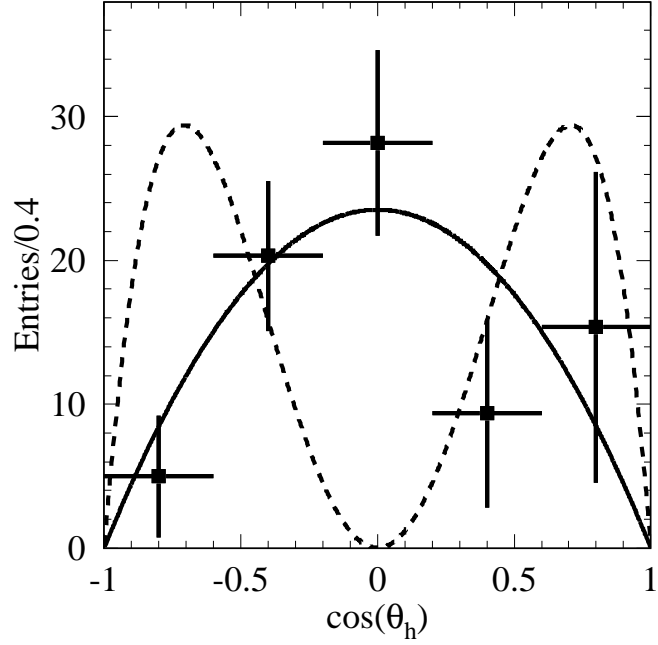}
 \end{center}

  \caption{Helicity distribution obtained from $m(D_s\gamma)$ fits in bins of $\cos(\theta_h)$
           for data (points) in comparison with the expectations for a $\DsFE$ spin
           $J=1$ (solid line) and $J=2$ (dashed line), respectively, after normalizing the predicted spectra
           to the data.
} \label{fig:helicity}
\end{figure}
The data 
exclude the
$J=2$ hypothesis ($\chi^2$/n.d.f.=36.4/4), while they are in good agreement
with the hypothesis of $J=1$ ($\chi^2$/n.d.f.=4.0/4). A  $J=0$ spin is ruled
out by parity and angular momentum conservation in the decay
$\DsFE \rightarrow D^+_s\gamma$.

\subsection{Observation of the $X(3872)$ in $B$ decays}

A new state, known as $X(3872)$, was recently discovered
by the Belle Collaboration~\cite{X-ref2} and confirmed by CDF~\cite{X-ref3}.
This state, reconstructed in the final state $\Jpsi\pip\pim$, 
is probably a charmonium candidate for the state $1^3D_2$ ($J^{PC}=2^{--}$) or $1^3D_3$ 
($J^{PC}=3^{--}$)~\cite{X-ref4},
but  could also be a molecule of $D$ and $D^*$ mesons~\cite{X-ref5}.

\babar\ recently confirmed the existence of this particle~\cite{X}.  
The analysis, based on 117 million \BBbar\ events,
reconstructs the $X(3872)$ in the final state 
$\Jpsi\pip\pim$ for $X(3872)$ produced in the decay $\Bm\to X(3872)\Km$.  
A total of $30.6\pm5.8$ events are reconstructed.
The mass of the $X(3872)$ is measured to be
$ m(X(3872))=(3873.4\pm1.4)\,\MeVcc, $
and the product branching fraction
$ \BF(\Bm\to X(3872)\Km)\times\BF(X(3872)\to\Jpsi\pip\pim)=(1.28\pm0.41)\times10^{-5} $
is obtained.

\subsection{Search for the $D^*_{sJ}(2632)^+$ meson}

A narrow state decaying to $\Dsp\eta$, named $D^*_{sJ}(2632)^+$, has been recently 
observed by the SELEX Collaboration~\cite{D3-ref1}. 
\babar\ has searched for this resonance in the final states $\Dsp\eta$, $\Dz\Kp$ and $\Dstp\KS$. 
The analysis, based on 125\,\fbinv, aims to reconstruct $\DsJ$ states produced in $\epem\to 
\ccbar$ events. 

Figure~\ref{D3-fig1+2+5} shows the abundant and clean samples of $\Dsp\to\Kp\Km\pip$ 
(left)
and $\eta\to\gamma\gamma$ (middle) used for this analysis.
\begin{figure}\centering
\includegraphics[width=5.2cm]{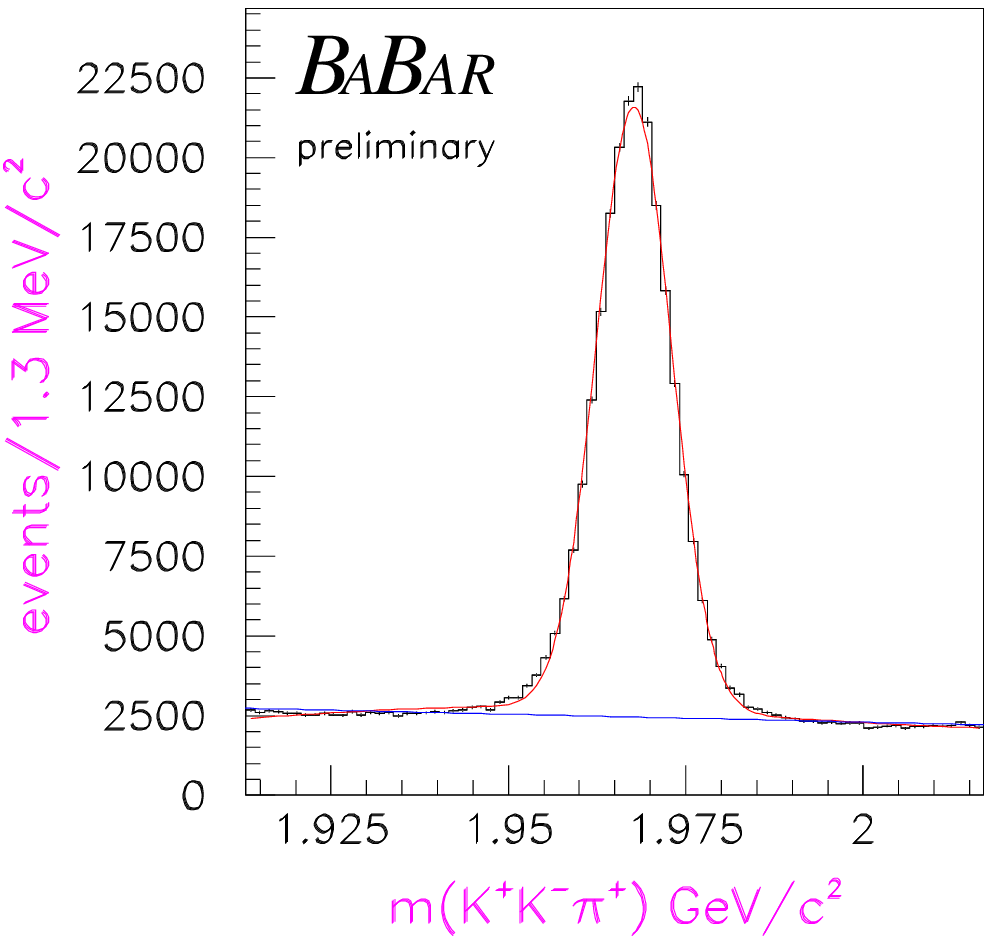}
\includegraphics[width=5.2cm]{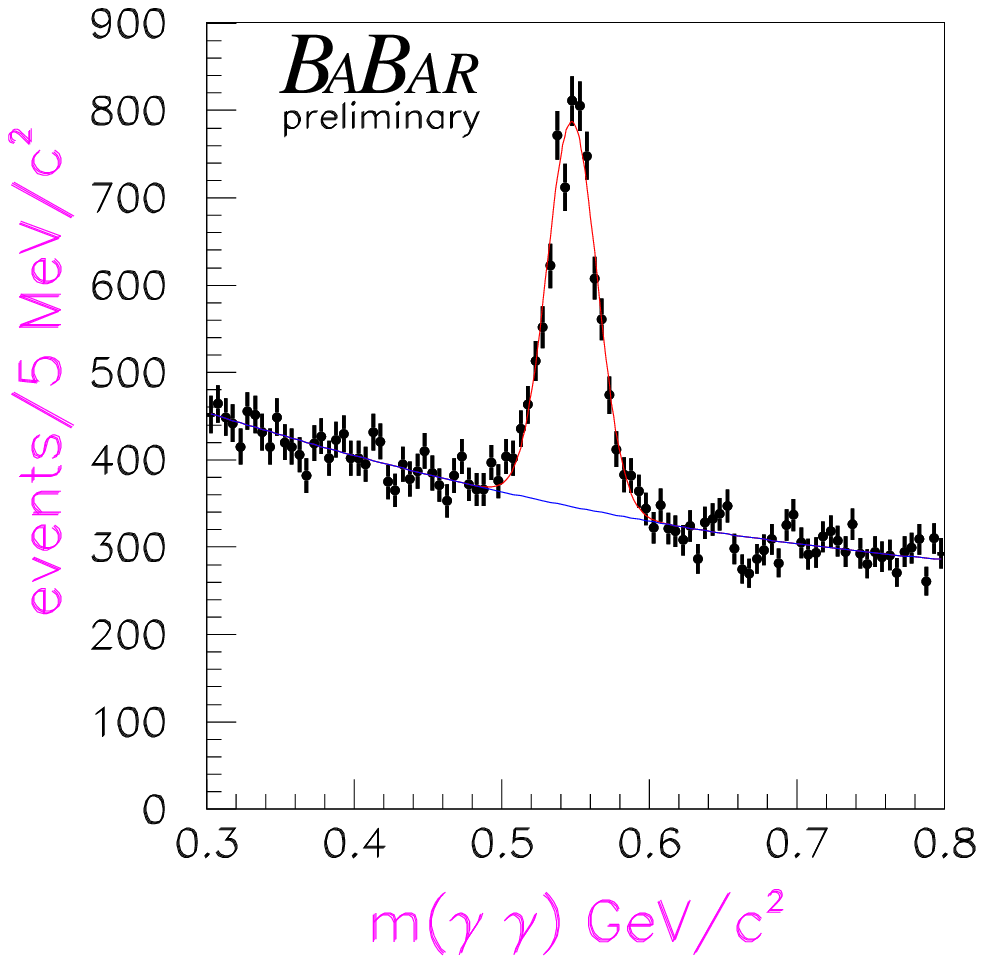}
\includegraphics[width=5.2cm]{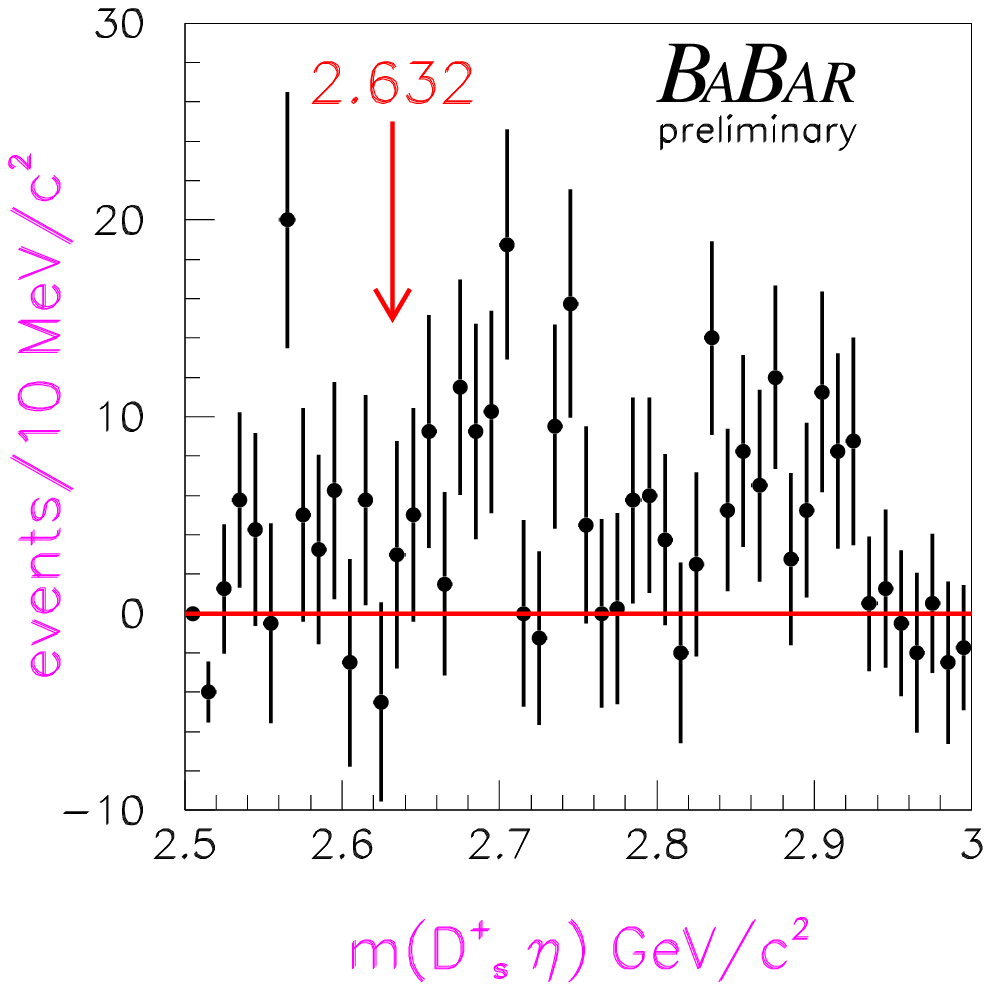}
\caption{$\Kp\Km\pip$ (left), $\gamma\gamma$ (middle),
  and $\Dsp\eta$ (right) invariant mass distributions in
  125\,\fbinv\ .}
\label{D3-fig1+2+5}
\end{figure}
The background subtracted invariant 
mass of the $\Dsp\eta$ system is shown in figure~\ref{D3-fig1+2+5} (right).
No signal is evident in the 
region around 2632\,\MeVcc. 
Figure~\ref{D3-fig8+10} shows the invariant mass of the systems $\Dz\Kp$ (left) and 
$\Dstp\KS$ (right).
\begin{figure}\centering
\includegraphics[width=6.5cm]{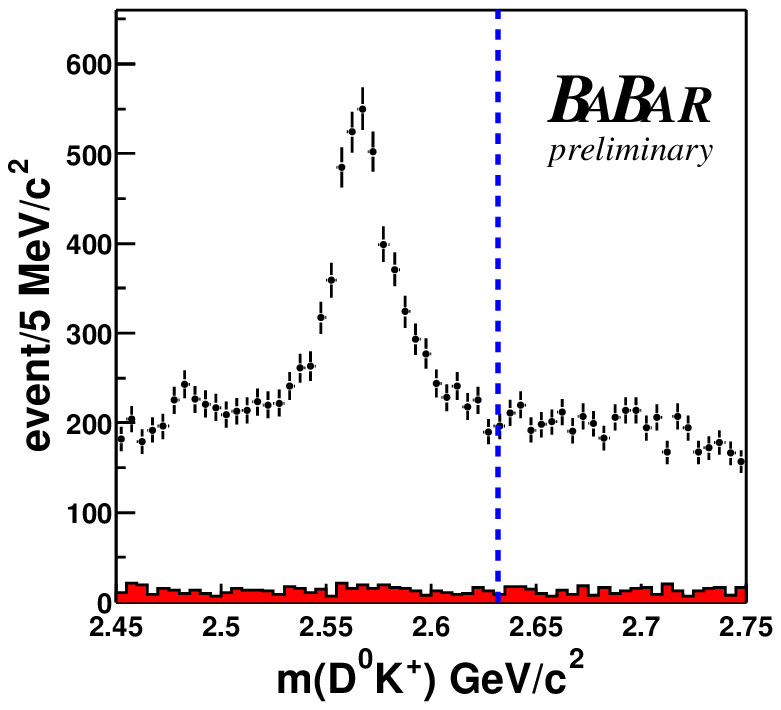}
\includegraphics[width=6.5cm]{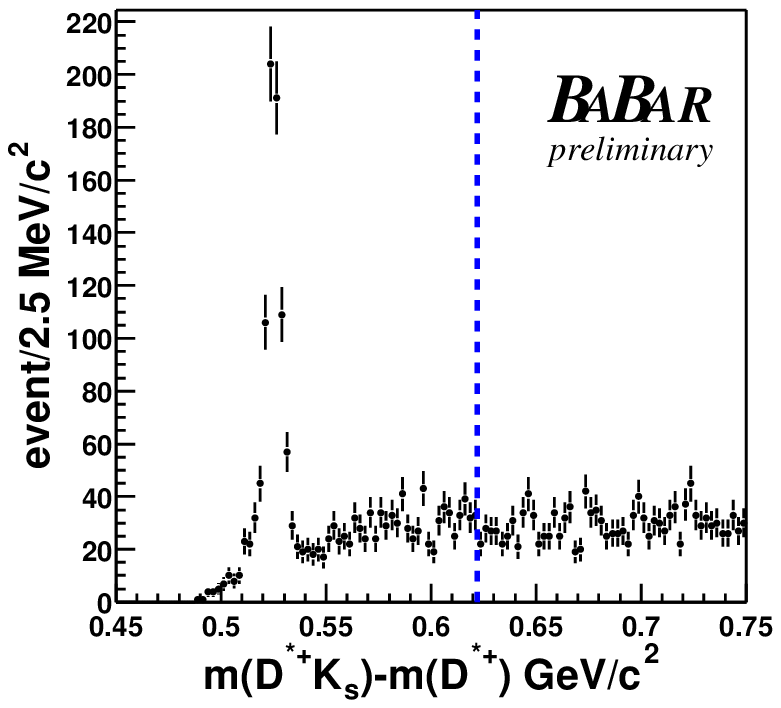}
\caption{$\Dz\Kp$ (left) and $\Dstp\KS$ (right) invariant mass
  distributions in 125\,\fbinv .}
\label{D3-fig8+10}
\end{figure}
These distributions show very clear signals for the decays 
$D_{s2}(2573)^+\to\Dz\Kp$ and $D_{s1}(2536)^+\to\Dstp\KS$, but no accumulation 
where the $D^*_{sJ}(2632)$ is expected.  

\section{Pentaquark searches at BABAR}

In the past two years, several experiments reported evidence for new baryon states that are 
compatible with being bound states of five quarks (pentaquarks). The most convincing evidence is for 
an exotic ($B=1$, $S=1$), light resonance called $\Theta^+(1540)$~\cite{pq-ref1,pq-ref2,pq-ref3,pq-ref4,pq-ref5,pq-ref6,pq-ref7}. This 
state has an unusually narrow width for a particle that could decay strongly. 
Recently, NA49 published evidence for  two heavier narrow states
($\Xi^{--}$ and  $\Xi^{0}$)~\cite{pq-ref8},  while H1 observed 
a narrow exotic charmed resonance ($\Theta_c^0$) with mass of 3099\,\MeVcc~\cite{pq-ref9}.  

Several theoretical models have been proposed to describe the pentaquark
structure~\cite{pq-ref11,pq-ref12,pq-ref13}.
The prediction is that the lowest mass states containing $u$, $d$ and $s$ quarks should 
occupy a spin-$\frac{1}{2}$ anti-decuplet and octet as illustrated in figure~\ref{pq-fig1}. 
\begin{figure}\centering
\includegraphics[width=10cm]{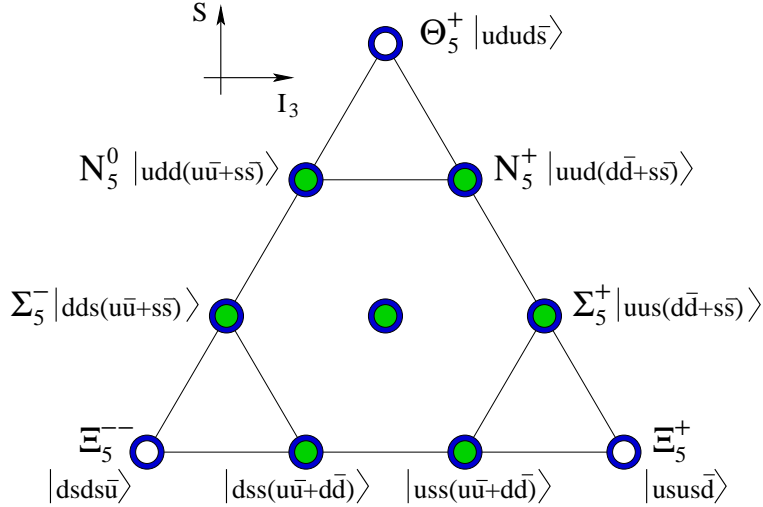}
\caption{Light pentaquark anti-decuplet and octet.}
\label{pq-fig1}
\end{figure}

Although experiments with baryon beams may have an advantage in producing these states, we 
know that $\epem$ collisions produce quite democratically all of the other particles.
It has been experimentally observed that for baryons with non-zero $B$, $C$ or strangeness, and/or orbital angular 
momentum, the production rates seem to depend only on mass and spin, and not on the quark content. 
If pentaquarks are produced similarly, their production rate should be as high as that for 
ordinary baryons of the same mass and spin, i.e.,
about $8\times10^{-4}$ for the $\Theta_5^+(1540)$ and  $4\times10^{-5}$
for the $\Theta_5^{--}$~\cite{pq-ref15}. 

It has been recently suggested that heavier pentaquarks 
such as the $\Theta^{*++}$
could be accessible at the B factories.  
The $\Theta^{*++}$ is a member of the baryon 27-plet
containing the quarks $uuud\overline{s}$ and has $I=1$ and $I_3=1$. 
It is predicted to decay into the final state $pK^+$, and to have a mass between
1.43 and 1.70\,\GeVcc\ with  a width of 37--80\,\MeVcc~\cite{pq2-ref6}. 

In this section I  discuss 
the results of several searches for light~\cite{pq} and 
heavy~\cite{pq2} pentaquarks recently published by  \babar .  
Light strange pentaquarks 
belonging to the decuplet and octet represented in figure~\ref{pq-fig1} 
are inclusively searched for in 123\,\fbinv. 
The analysis focuses on final states containing strange particles and protons that are easily 
identifiable in the detector. The search for the $\Theta^{*++}$ follows a very different 
approach, involving the exclusive reconstruction of $\Bp$ mesons decaying into $p\pbar K$ final 
states, as suggested in~\cite{pq2-ref5}. This analysis is performed with 81\,\fbinv.

\subsection{Search for $\Theta_5^+(1540)\to p\KS$}

This analysis aims to reconstruct decays of the 
$\Theta_5^+(1540)$  into the  $p\KS$ final state. The $\KS$ candidates are 
reconstructed in their decay into two charged pions, with selection criteria designed to avoid a bias toward 
any specific production mechanism, while keeping a high reconstruction efficiency.
The $\KS$ candidates are then combined with protons from the interaction point
to form the $\Theta_5^+(1540)$ candidates.

The invariant mass distribution of the $p\KS$ pairs is shown in figure~\ref{pq-fig2}.
\begin{figure}\centering
\includegraphics[width=8cm]{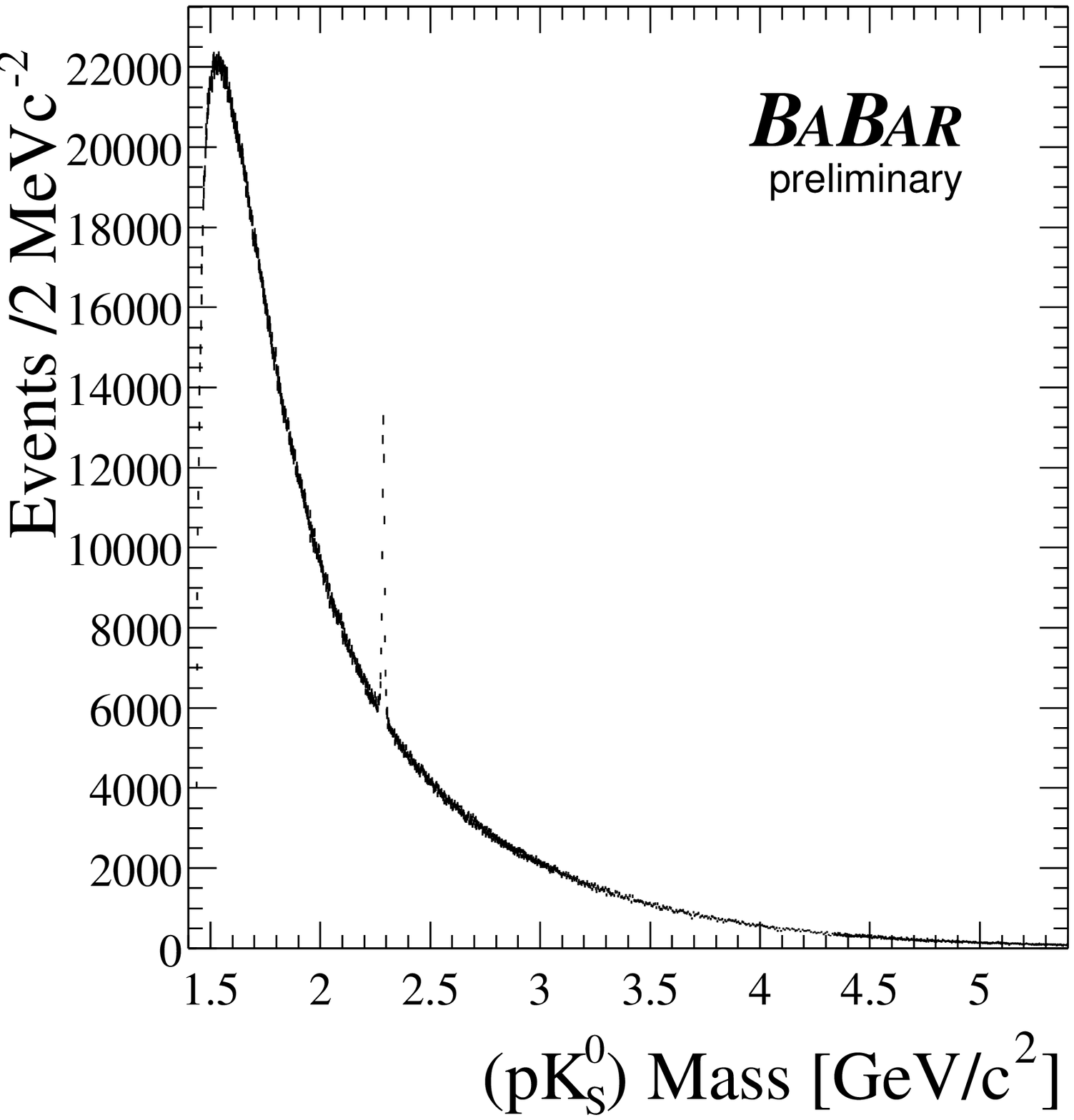}
\includegraphics[width=8cm]{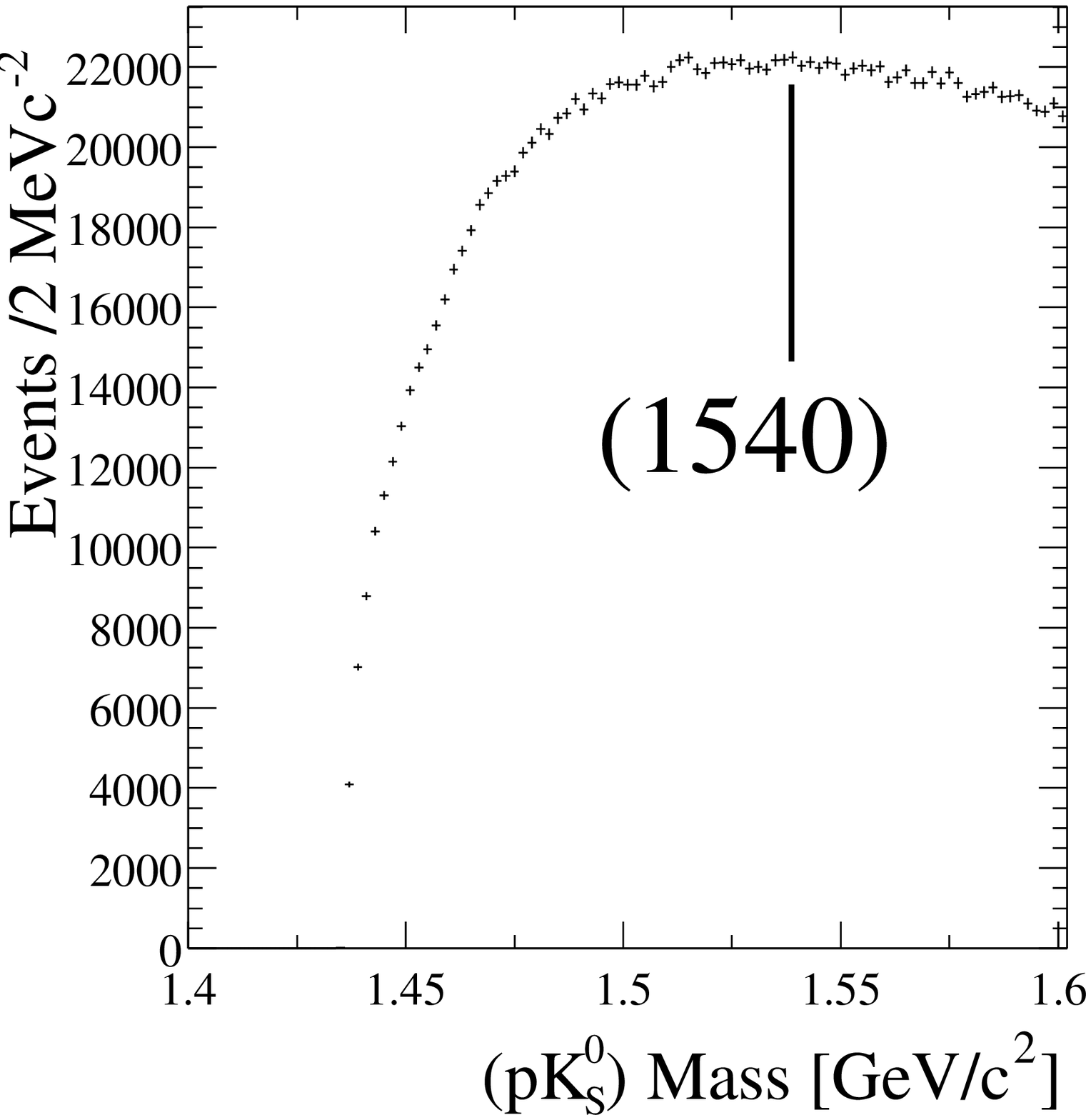}
\caption{$p\KS$ invariant mass distributions in the $\Theta_5^+(1540)$ search.}
\label{pq-fig2}
\end{figure}
The clear peak around  2285\,\MeVcc\ visible on the left plot is due to the decay 
$\Lambda_c^+\to p\KS$.
The abundant signal and the narrow peak 
demonstrate our sensitivity to the presence of a narrow resonance.
The left plot of figure~\ref{pq-fig2} is a magnified view of 
 the region where the $\Theta_5^+(1540)$ is expected.
No signal is observed. 
Null results are obtained also when the search is performed 
separately for ten momentum bins uniformly distributed between 
zero and 5\,\GeVc . 

The null results are used to set a limit on the differential production cross section of 
$\Theta_5^+(1540)$ in $\epem$ interaction. In the calculation the mass is assumed to be
1540\,\MeVcc. Since the natural width of this particle is unknown, two hypotheses have been 
considered: $\Gamma=8\,\MeVcc$, corresponding to the experimental upper limit, and 
$\Gamma=1\,\MeVcc$, corresponding to a very narrow state. The 95\% C.L.\ upper limit on the 
differential production cross section of the $\Theta_5^+(1540)$	is shown in
figure~\ref{pq-fig14}. 
\begin{figure}\centering
\includegraphics[width=7.5cm]{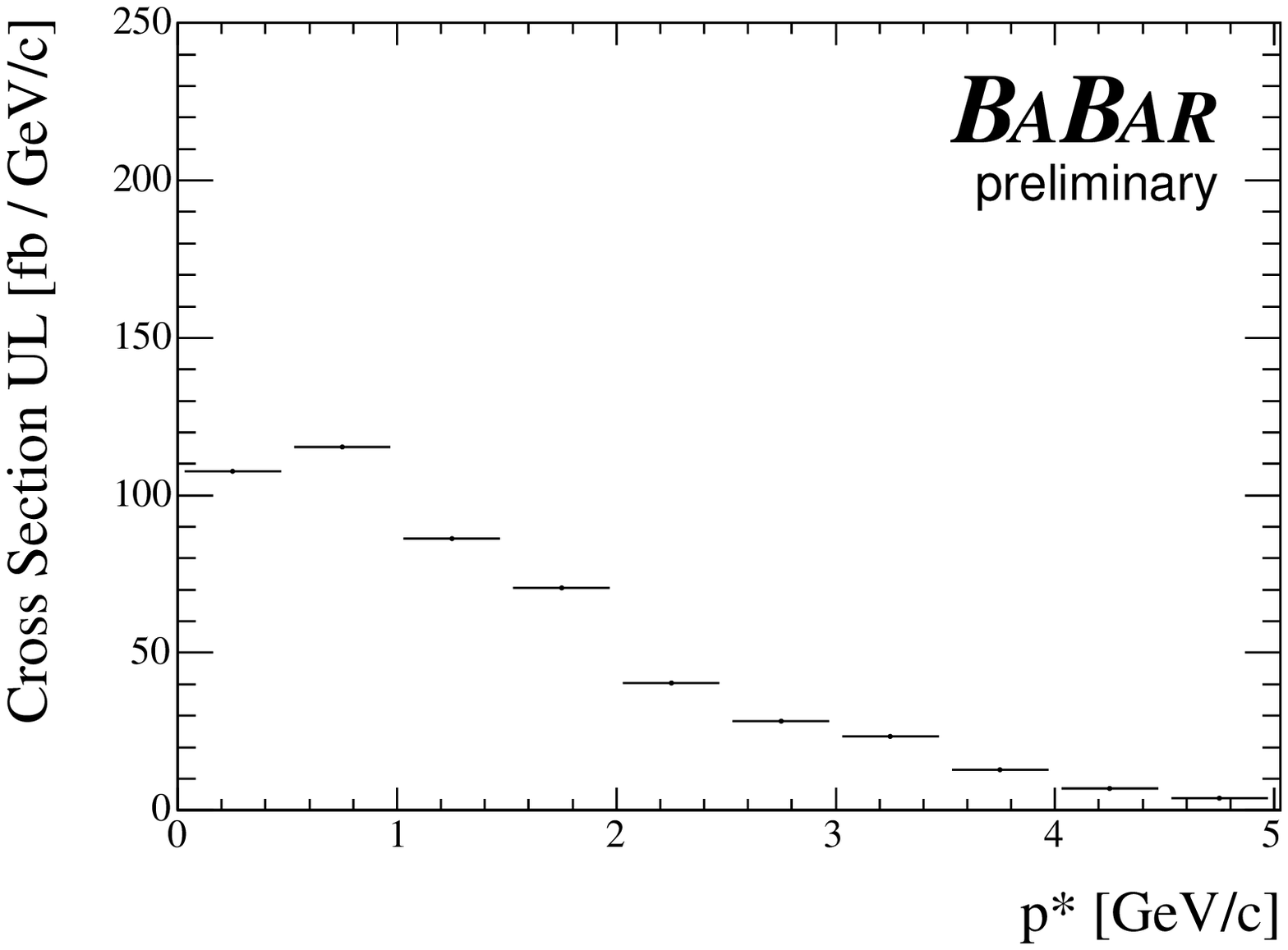}
\includegraphics[width=7.5cm]{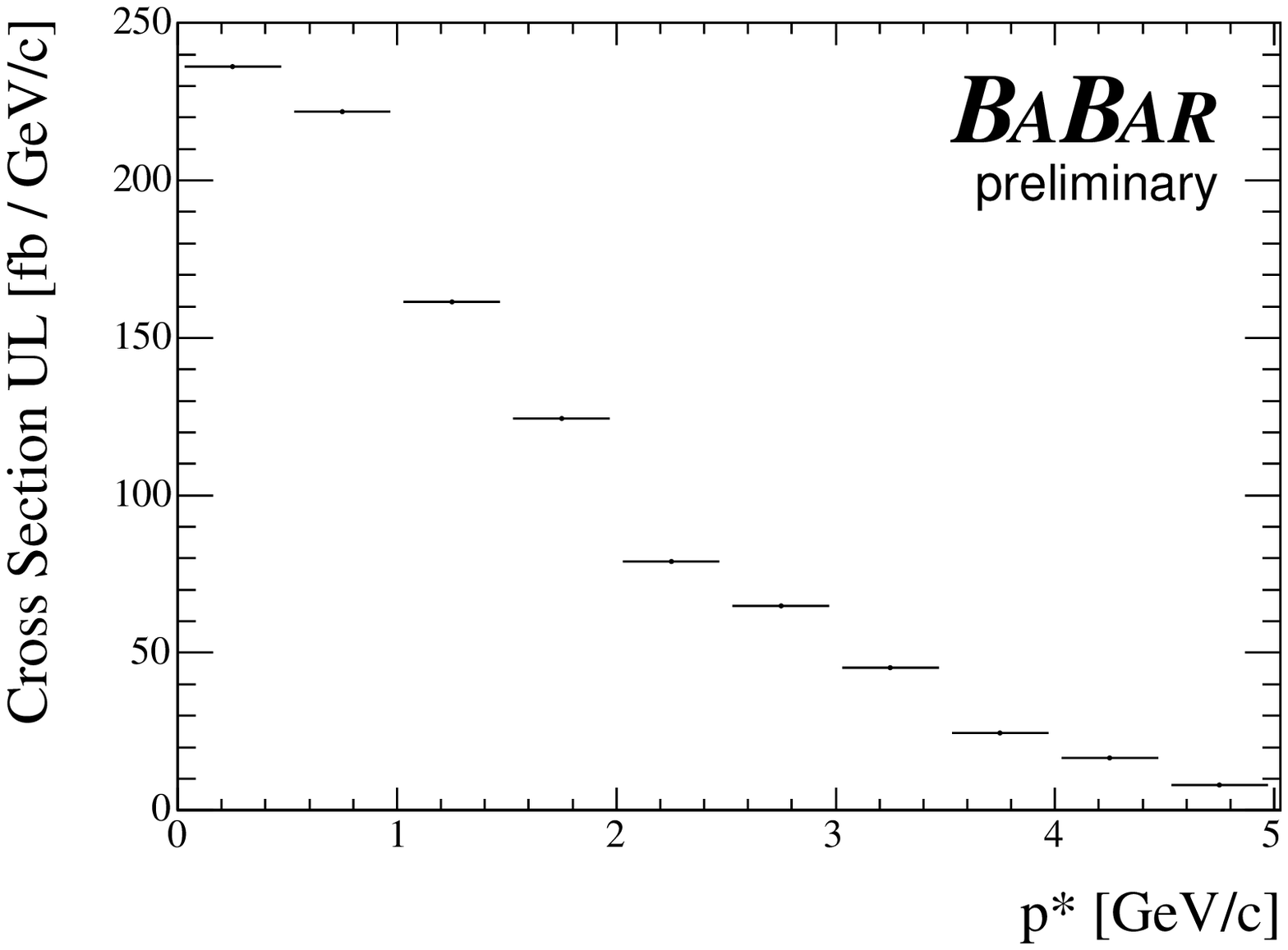}
\caption{95\% C.L.\ upper limit of the differential production cross section of
  $\epem\to\Theta_5^+(1540)X$, assuming $\Gamma=8\,\MeVcc$ (left) and $1\,\MeVcc$ (right).}
\label{pq-fig14}
\end{figure}
The branching fraction $\BF(\Theta_5^+(1540)\to p\KS)$ is assumed to 
be $1/4$.  Taking the upper limit width, the 95\% C.L.\ upper limit on the total production rate is 
calculated to be $1.1\times10^{-4}$ per $\qqbar$ event, roughly a factor of eight below the values 
measured for ordinary baryons of similar mass as shown in figure~\ref{pq-fig17}. 
\begin{figure}\centering
\includegraphics[width=10cm]{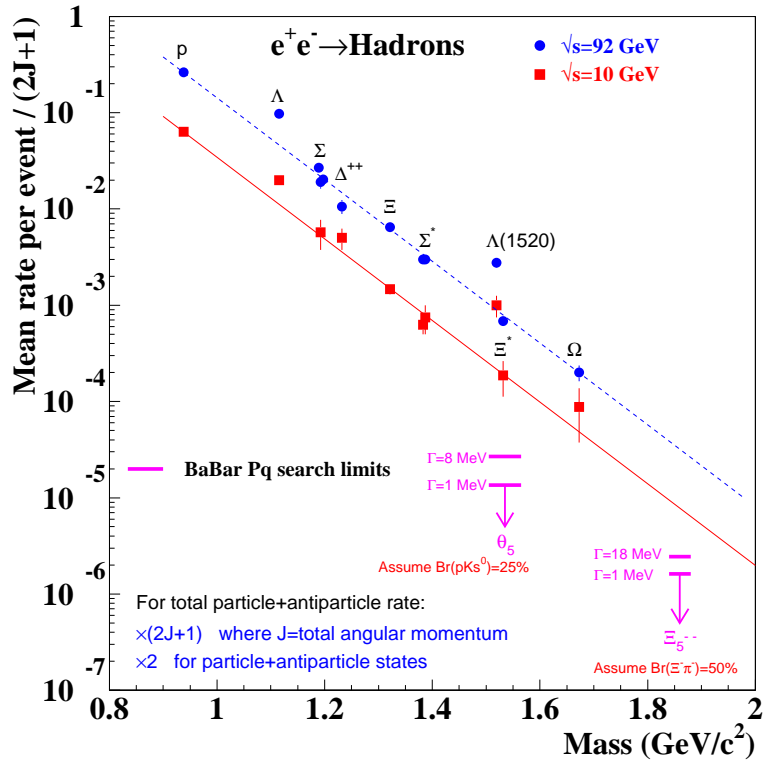}
\caption{%
Baryon production rates in \epem\ interactions \cite{pq-ref15} measured at the $Z^0$ pole (circles) and
$\Upsilon(4S)$ (squares).
The vertical axis accounts for number of spin and particle+antiparticle states. 
The lines are chosen to guide the eye. The arrows indicate the upper limits on 
the $\Theta_5^+$ and $\Xi_5^{--}$.}
\label{pq-fig17}
\end{figure}

\subsection{Search for $\Xi_5^{--}\to\Xi^-\pi^-$ and $\Xi_5^{0}\to\Xi^-\pi^+$}

The $\Xi_5^{--}$ and $\Xi_5^{0}$ states were searched for in decays involving $\Xi^-$ baryon 
and a charged pion. The $\Xi^-$ is reconstructed in its decay $\Xi^-\to\Lambda^0\pi^-$ with 
$\Lambda^0\to p\pi^-$. As for the $\Theta_5^+(1540)$ search, all selection criteria used in the 
analysis are designed to maximize the efficiency and avoid a bias toward any specific 
production mechanism.

The invariant mass distributions for the $\Xi^-\pi^-$ and $\Xi^-\pi^+$ pairs are illustrated in 
figure~\ref{pq-fig4+5}.
\begin{figure}\centering
\includegraphics[width=8cm]{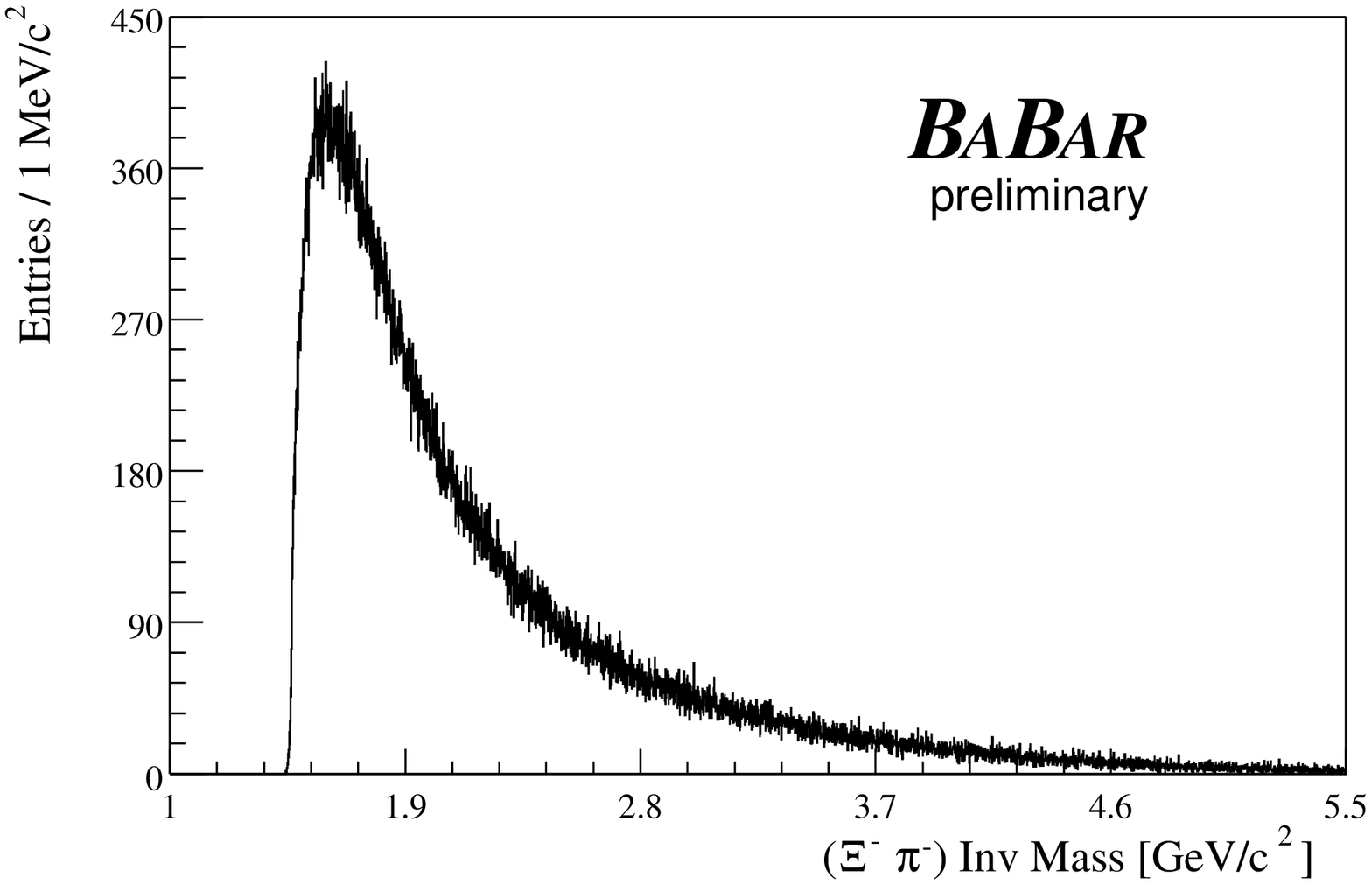}
\includegraphics[width=8cm]{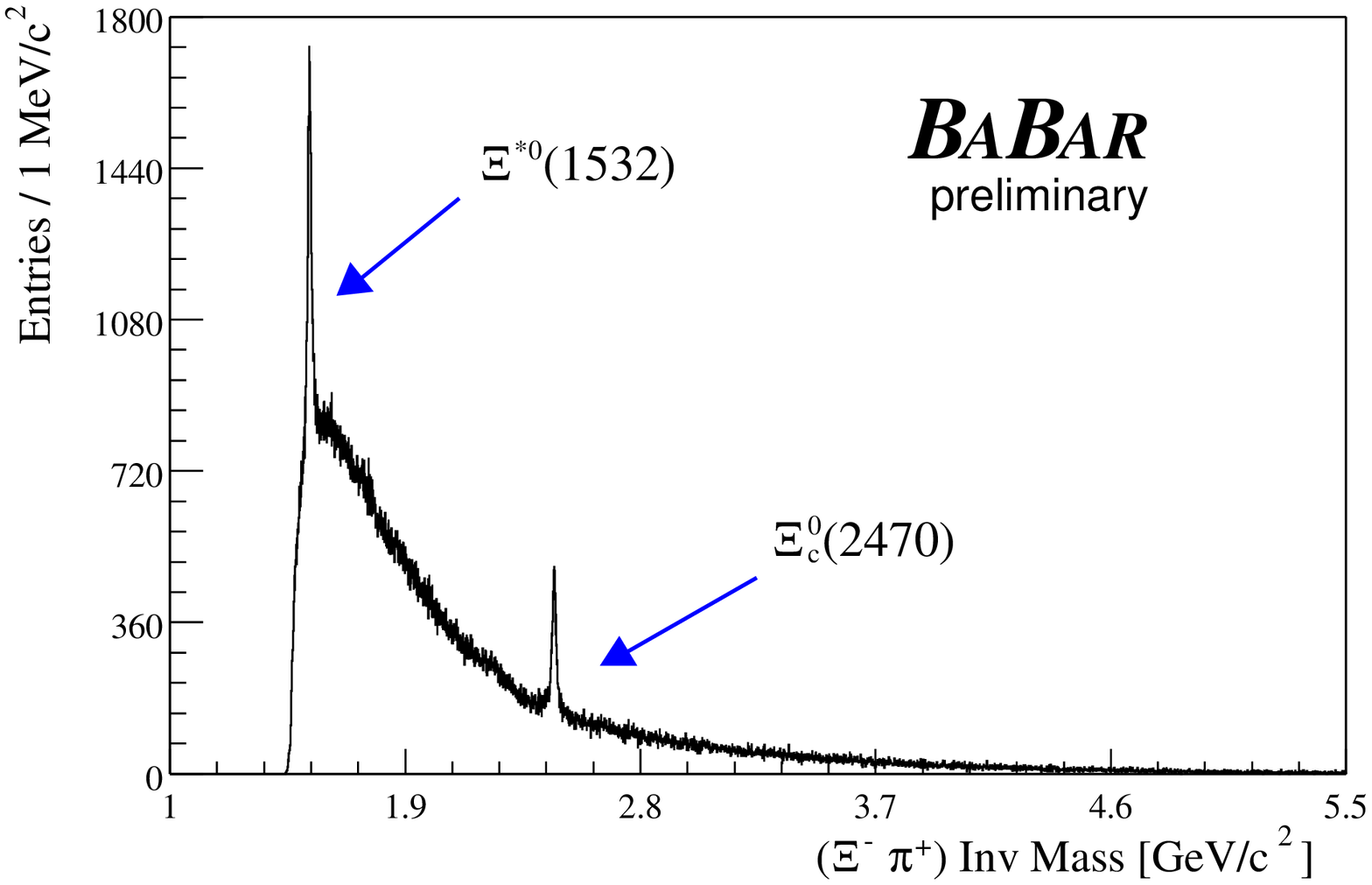}
\caption{$\Xi^-\pi^-$ (left) and $\Xi^-\pi^+$ (right) invariant mass distributions
  in the $\Xi_5^{--}$ and $\Xi_5^{0}$ searches.}
\label{pq-fig4+5}
\end{figure}
No resonance is visible in the $\Xi^-\pi^-$ 
invariant mass distribution. 
Two peaks are instead visible in the  $\Xi^-\pi^+$ plot, 
corresponding to the decays of $\Xi^{*0}(1532)$ and $\Xi^0_c(2470)$, but 
no structure is visible in the region where the $\Xi_5^{0}$ is expected. 

The absence of a $\Xi_5^{--}$ signal is reflected as an upper limit at the 
95\% C.L. on the differential 
production cross section shown in figure~\ref{pq-fig15} as a function of $p^*$, the momentum 
in the $\epem$ center of mass frame. 
\begin{figure}\centering
\includegraphics[width=9cm]{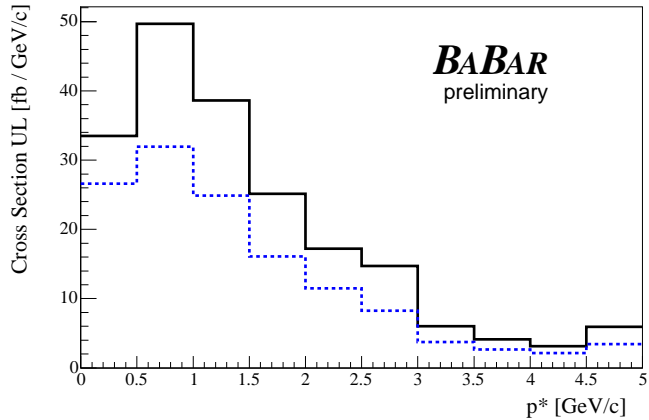}
\caption{95\% C.L.\ upper limit of the differential production cross section of
  $\epem\to\Xi_5^{--}X$, assuming $\Gamma=18\,\MeVcc$ (solid line) and $1\,\MeVcc$ (dotted line).}
\label{pq-fig15}
\end{figure}
The branching fraction $\BF(\Xi_5^{--}\to\Xi^-\pi^-)$ is assumed to 
be $1/2$.  Assuming a width of $18\,\MeVcc$, the 95\% C.L.\ upper limit on the total production rate is 
calculated to be $1.0\times10^{-5}$ per $\qqbar$ event, roughly a factor of two below  the values 
measured for ordinary baryon of similar mass as illustrated in figure~\ref{pq-fig17}. 

\subsection{Inclusive searches for other strange pentaquarks}

\babar\ also performs a more inclusive search for other strange pentaquarks 
decaying into the final states $\Lambda^0 K$ and $\Sigma^0 K$, where $K$ is either a charged or a 
neutral kaon and $\Sigma^0 \to \Lambda^0 \gamma$ and $\Lambda^0 \to p\pi^-$. 
These final states give access to the pentaquarks $\Xi_5^-$,  $\Xi_5^0$,  $N_5^0$ and $N_5^+$ 
introduced in figure \ref{pq-fig1}.  Inclusive searches are also performed for 
$\Sigma_5^+\to p\KS$. 
No pentaquark signal is observed in any of these inclusive searches. 

\subsection{Search for the $\Theta^{*++}$ pentaquark in exclusive B decays}

In this analysis, the $\Bp$ mesons are exclusively reconstructed in the $p\pbar\Kp$ final 
state. The purity of the signal is increased by cuts on the invariant mass and energy of the $B$
candidate, and by the use of topological variables that distinguish continuum
from $\BBbar$ events.
Of the 212 candidates reconstructed in this analysis, $188\pm17$ are estimated to be 
real $\Bp$ decays, including $68\pm10$ containing charmonium decays to $p\pbar$.

The $p\Kp$ invariant mass distribution of the selected $B$ candidates is analyzed 
searching for a pentaquark signal up to 2.36\,\GeVcc. No events are observed below 
1.85\,\GeVcc, and no structure is visible in the entire area studied.
The 95\% C.L.\ upper limit for the branching fraction of $\Bp\to\Theta^{*++}(pK^+)\pbar$
is measured to be
$1.5\times10^{-7}$ for $1.43<m(\Theta^{*++})<1.85\,\GeVcc$,
$2.4\times10^{-7}$ for $1.85<m(\Theta^{*++})<2.00\,\GeVcc$, and
$3.3\times10^{-7}$ for $2.00<m(\Theta^{*++})<2.36\,\GeVcc$.

\section{Summary}

Many new results in the field of hadronic physics were recently published by the \babar\ Collaboration.

The discovery of the $\DsTT$ narrow state revived 
the interest in charm spectroscopy. The mass of the $\DsTT$ and of its heavier twin, the 
$\DsFE$, are now measured with high precision in many different decay modes. In 
addition, the  measurement of other properties, such as branching fractions and 
angular distributions, are now accessible, thanks to the pure samples of $\DsJ$ mesons 
reconstructed in $B$ decays. 

The narrow state $X(3872)$ has recently been observed by \babar, in agreement with 
previous results from Belle.
The $\DsJ(2632)$ state reported by SELEX, on the other hand, has not been confirmed.  

\babar\ also searches for pentaquark production in $\epem$ interactions. Various states are  
investigated, and particular attention is given to the searches for the 
$\Theta_5^+(1540)$ and the $\Xi^{--}(1860)$. 
The null results reflect limits on the rates of the production of such 
states in $\epem$ events, well below the rates measured for ordinary baryons
of similar masses. 

\section*{Acknowledgments}
It is a pleasure to thank my \babar\ colleagues P.~Burchat, V.~Halyo, V.~Luth, M.~Morii and R.~Yamamoto for helpful 
comments. I would also like to thank the organizers of the APS Topical Group on Hadronic Physics 
for creating such a pleasant and productive environment. 
This work was supported by the DOE contracts DE-FC02-94ER40818 and DE-AC02-76SF00515.  


\end{document}